\documentclass[sigconf, manuscript]{acmart}
\usepackage{subcaption}
\usepackage[inline]{enumitem}
\usepackage{soul}
\usepackage{fontawesome}

\setcopyright{acmcopyright}
\copyrightyear{2020}
\acmYear{2020}
\acmDOI{10.1145/1122445.1122456}

\acmConference[CHI '21]{CHI '21: ACM Computer Human Interaction}{May
  8--13, 2021}{Yokohama, Japan}%
\acmBooktitle{CHI '21: ACM
  Computer Human Interaction, May 8--13, 2021, Yokohama, Japan}%
\acmPrice{15.00} \acmISBN{978-1-4503-XXXX-X/18/06} 

\begin{document}

\title{Proxemics and Social Interactions in an
  Instrumented Virtual Reality Workshop}



\author{Julie R. Williamson}
\affiliation{%
 \institution{University of Glasgow}
 \city{Glasgow}
 \country{Scotland}}
\orcid{0000-0002-1816-1393} 
\email{julie.williamson@glasgow.ac.uk}

\author{Jie Li}
\affiliation{%
 \institution{CWI}
 \city{Amsterdam}
 \country{Netherlands}}
\orcid{0000-0002-6791-104X} 
\email{jie.li@cwi.nl}

\author{Vinoba Vinayagamoorthy}
\affiliation{%
 \institution{BBC R\&D}
 \city{London}
 \country{England}}
\orcid{0000-0002-3301-9723} 
\email{vinoba.vinayagamoorthy@bbc.co.uk}

\author{David A. Shamma}
\affiliation{%
 \institution{CWI}
 \city{Amsterdam}
 \country{Netherlands}}
\orcid{0000-0003-2399-9374} 
\email{aymans@acm.org}

\author{Pablo Cesar}
\affiliation{%
 \institution{CWI}
 \city{Amsterdam}
 \country{Netherlands}}
\orcid{0000-0003-1752-6837} 
\email{p.s.cesar@cwi.nl}

\begin{abstract}
  Virtual environments (VEs) can create collaborative and social
  spaces, which are increasingly important in the face of remote work
  and travel reduction. Recent advances, such as more open and widely
  available platforms, create new possibilities to observe and analyse
  interaction in VEs.  Using a custom instrumented build of Mozilla
  Hubs to measure position and orientation, we conducted an academic
  workshop to facilitate a range of typical workshop activities. We
  analysed social interactions during a keynote, small group
  breakouts, and informal networking/hallway conversations. Our
  mixed-methods approach combined environment logging, observations,
  and semi-structured interviews.  The results demonstrate how small
  and large spaces influenced group formation, shared attention, and
  personal space, where smaller rooms facilitated more cohesive groups
  while larger rooms made small group formation challenging but
  personal space more flexible. Beyond our findings, we show how the
  combination of data and insights can fuel collaborative spaces'
  design and deliver more effective virtual workshops.
\end{abstract}




\begin{teaserfigure}
  \centering
  \includegraphics[width=0.75\textwidth]{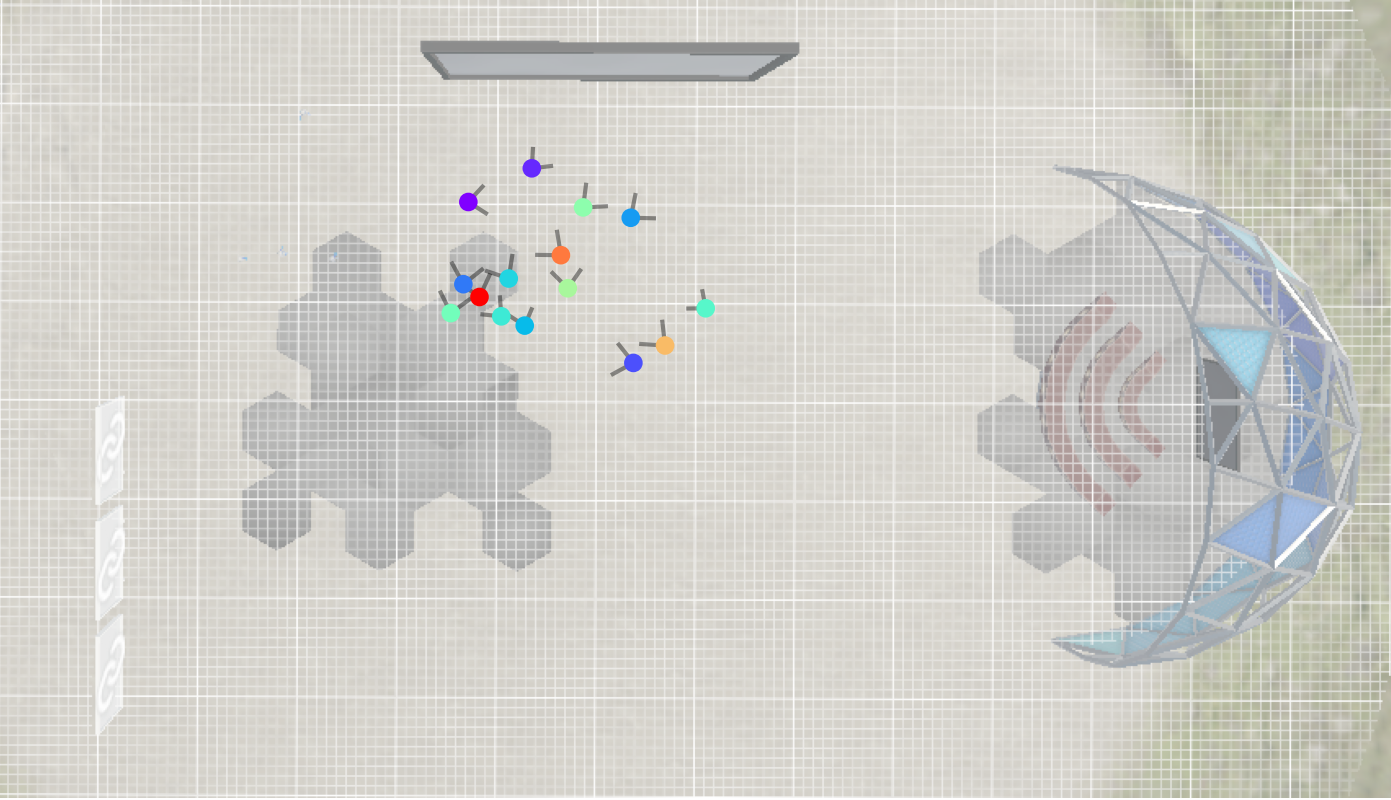}
  \caption{\label{fig:mainteaser}An overhead view of the main room showing 
  participants locations and viewports while they attend the workshop.}
  \Description{Top down view of outdoor virtual space.  A large
    display runs along the top wall.  A dome amphitheatre is along the
    right wall.  A floor texture of hexagons if off-centre to the
    left. A group of workshop participants stands in front of the large display.}
\end{teaserfigure}

\maketitle

\section{Introduction}
Virtual Environments (VEs) are not a replacement for face-to-face
interaction, but VEs can provide solutions in the face of increasing
pressures to reduce travel and work remotely. In this paper, we
explore how virtual academic workshops can be effective at enabling
social group activities like keynote talks, small group discussions,
and informal networking in a VE\@. Taking advantage of customisation
in the open source VE Mozilla
Hubs\footnote{\href{https://hubs.mozilla.com/}{Mozilla Hubs} is an
  open source platform for hosting virtual events.}, we built a custom
environment instrumented with position and movement logging. We
designed and evaluated how the workshop programme, the virtual spaces,
and hosting approaches contributed to a successful event.

Current VEs vary widely in affordances, fidelity, scale, and
accessibility.  VEs like Mozilla Hubs, Gather
Town\footnote{\href{https://gather.town/}{Gather Town} combines 2D
  maps with video conferencing.}, and
VRChat\footnote{\href{https://www.vrchat.com/}{VRChat} is an online
  massively multiplayer social environment.} all create social VEs,
but result in dramatically different experiences. Mozilla Hubs and
VRChat provide fully 3D environments that can be experienced on a
desktop or using a head-mounted display (HMD). Gather Town uses a 2D
map, but incorporates video conferencing for groups to chat. VRChat is
a massively online multiplayer environment, averaging over 10,000
users daily.  In contrast, Mozilla Hubs and Gather Town support a
maximum of 25 and 50 users respectively, although premium Gather Town
rooms can host up to 500. VRChat requires a relatively high
specification PC suitable for gaming, but Mozilla Hubs and Gather Town
run in most standard browsers. Accessibility in VEs presents an
on-going challenge~\cite{Trewin2008AccessibilityWorlds}, and open
source VEs like Mozilla Hubs create the best opportunities to address
accessibility requirements.  This is especially important for
scholarly community events which must prioritise accessibility and
inclusion.  Organisers designing virtual events must balance complex
issues and often completing needs~\cite{10.1145/3406108}, making it
challenging to choose the right tools for the experience they are
trying to create.  For an academic workshop, our priorities were to
facilitate social interaction in small groups, create opportunities to
network, and facilitate discussion on open-ended topics.  As
researchers in social virtual reality (VR), we wanted to organise an
event to discuss the open issues \textit{of} social VR \textit{in}
social VR\@.  Although we initially intended this as a hybrid event
with physical and virtual spaces, the pandemic of 2020 meant we had to
rapidly adapt to a fully virtual event.

In this paper, we present research on running an academic workshop in
Mozilla Hubs.  The workshop programme included a keynote talk,
individual pitches, small group breakouts, and information
breaks/networking opportunities. We designed the virtual environments
to support these activities using a large outdoor environment for
collective activities (see Figure~\ref{fig:mainteaser}) and breakout rooms for smaller groups.  We
customised a Mozilla Hubs Cloud instance to log user positions and
orientations, generating a proxemic dataset from the workshop event.
Combining a proxemic dataset with observations and post-workshop
interviews, we explore whether proxemic interactions in a VE can be analysed using metaphors from physical settings, what is the relationship between group activity and VEs,
and what social cues are available or needed? To answer these
questions, we used an open source VE instrumented to support
quantitative analysis of position, orientation,
framerate, and user input as social signals, and release the code
for collection, analysis,
and captured dataset (see Appendix~\ref{app:dataset}).  We demonstrate
the use of quantitative analysis
for measuring proxemics, interactions, and personal space, inspired by
interaction in physical settings. We combine this with a mixed methods
analysis of group formation and social cues to see how the environment
design affects small and large groups collaborating in a VE\@.
Finally, we conclude with a discussion of open challenges for social
group interaction in academic workshops in VEs, including dynamics
like expressing oneself effectively, knowing where to stand, and
interrupting existing
conversations. 

\section{Background}
There is a substantial body of work on socialising and collaborating
in VEs that informed our approach.  We discuss the existing methods
for evaluations in VEs, and how our instrumented approach adds
insights to these methods.  We also draw from literature on proxemics
and personal space, discussing how these concepts from physical
environments translate to the virtual.  Finally, we discuss research
on group formation and social regulation in VEs and their use on
informing the design of social activities.

\subsection{Methods for Analysing Behaviour in Virtual Environments}
Previous findings on proxemics, spatial positioning, and group
interaction in VEs are primarily supported by observational data and
questionnaires.  Metrics based on standardised questionnaires inform
us about the perceived presence and
immersion~\cite{Slater1994DepthEnvironments,
  Witmer1998MeasuringQuestionnaire}. More recently, an extension to
these questionnaires aims at measuring the quality of the interaction
of participants in a social VR
experience~\cite{Li2019MeasuringReality}. These methods require post
self-reflection from the users and are not suitable for capturing
actual movement and behaviour of users. Ethnographic methods based on
observations have been used to better understand user behaviour. For
example, observational notes can be used to analyse
behaviour~\cite{Le2020EnhancingEnvironments}. Additional examples
include the study of user behaviour in immersive games like
`there'~\cite{Brown2004CSCWEnvironment} or World of
Warcraft~\cite{Ducheneaut2006Games}, and in large virtual environments
like Second Life~\cite{Yee2007TheEnvironments}. Audio and voice have
also been captured for analysing interaction in
VEs~\cite{Bowers1996TalkEnvironments, Wadley2015VoicePlay,
  Carter2015VoiceInteraction}.

There are limited examples of more systematic instrumentation of VEs for analysing user position and movement. For example, Friedman et
al.~\cite{Friedman2007SpatialLife} explores the spatial social
behaviour in Second Life based on software bots that wander around and
collect data, while Ranathunga et
al.~\cite{Ranathunga2012IdentifyingEvents} reports an infrastructure
for identifying events and happenings in Second Life. More recently,
Le at al.~\cite{Le2020EnhancingEnvironments} perform quantitative
analysis of the behaviour of participants in a virtual conference
based on the logs from a voice and chat application (Discord)
integrated in a virtual environment (Mozilla Hubs).

VEs are growing in popularity, influencing training, collaboration,
social experiences~\cite{Slater2016EnhancingOurLives}. Previous work
on scientific events has explored large-scale
gatherings~\cite{Erickson2011SynchronousInteractionHundreds} and the
affordances of the existing
technology~\cite{Shirmohammad2012Mixingvirtualphysicalparticipation,
  Neustaedter2016RemoteTelepresenceAttendance}, in some cases with
limited instrumentation of the space or the auxiliary communication
channels~\cite{Le2020EnhancingEnvironments}.  In contrast, our work is
focused on understanding the individual and group behaviour of
participants attending a small and focused scientific event in a
VE\@. Our method is inspired by interaction in physical settings, for
example using cameras~\cite{CabreraQuiros2020GesturesID}, mobile
devices~\cite{Chin2012, Guo2016}, and wearable
sensors~\cite{Kim2008MeetingMediator} to capture high resolution data
on position and orientation.  We also completed observations and
post-workshop interviews to gather qualitative insights.  Thus, our
approach uses mixed-methods, drawing from both qualitative and
quantitative techniques to analyse social interaction in a VE\@.

\subsection{Proximity and Personal Space}
The flow and changes in interpersonal distances between individuals in
a shared space is an integral part of nonverbal communication. Neither
the intention nor awareness of specific behavioural cues is necessary
and yet these cues dictate part of the conversation. In fact, it adds
to the richness and versatility of the
conversation~\cite{Argyle1988}. Hall's theoretical model of proxemics,
largely based on middle class American adults, focused on four
concentric zones to define different levels of social intimacy which
`informed' the interpersonal distance maintained in between
people~\cite{Hall1969ThePrivate}. These `circles' were
\textit{intimate} (<0.45m), \textit{personal} (0.45m-1.2m),
\textit{social} (1.2m-3.6m) and \textit{public} (>3.6m). However,
there are many factors which affect the proximal relationship between
people including age, culture, environmental context,
interpersonal relationship and emotional state of the
individuals~\cite{Argyle1979, Duck1998, Argyle1988,
  Scheflen1972BodyLA, Hayduk1983PersonalStand.}. Proximal behaviour is
also interconnected with other modalities of expression such as gaze
behaviour. Argyle and Dean's~\cite{Argyle1965} theory of equilibrium
posits that appropriate interpersonal distance is maintained between
individuals by balancing increasing proximity with reduced eye
contact. The lower limits of the distance is determined by physical
contact whereas the upper limit is defined by factor of visibility and
audibility~\cite{Hall1969ThePrivate}.

VEs are often crafted as representations of physical
spaces. However,
it is unclear how much of our understanding of physical spaces
transfers to analysing virtual spaces.  Users may enter a shared
virtual space with physical metaphors in mind, but in the absence of
physical cues and constraints do these metaphors transfer across?
Hecht et al.~\cite{Hecht2019TheSpace} reported that the shape of
personal space in the real world was remarkably close to a circular
zone with a radius of about one meter and this was similar to the
personal distance participants maintained to an avatar in a virtual
space. Bailenson et al~\cite{Bailenson2001Equilibrium} tested the
equilibrium theory in an immersive VE using two objective
measures---minimum distance participants maintained while approaching
a virtual male agent and invasion duration within the agents' intimate
space. They reported that participants avoided direct contact with the
agent and respected the personal space of the agent (relative to the
control condition: a cylinder). There were also gender differences
with female participants responding to mutual gaze behaviour with the
agent more than male participants. Bailenson et
al.~\cite{Bailenson2003InterpersonalEnvironments} extended their work
to include head gestures, female virtual agents and perceived
agency. Again, participants maintained greater personal distance with
the agent that engaged in mutual gaze in addition to getting out of
the way when the virtual agent approached the participants. This
effect was also noted by Llobera et
al.~\cite{Llobera2010ProxemicsEnvironment} where they noted that the
shorter the distance between virtual agents and the participant, the
greater the physiological arousal as indicated through electrodermal
activity. Wilcox et al.~\cite{Wilcox2006PersonalReality} further
reported emotional discomfort in situations of close encounters, both
through self-reported means and increased skin conductance. Yee et
al.~\cite{Yee2007TheEnvironments} conducted a observational study in
Second Life to explore if social norms of gender, interpersonal
distance and gaze behaviour transferred to VEs even though
participants used a keyboard and mouse as opposed to their
bodies. They found support for the equilibrium theory through
avoidance of collision---a result reported by Friedman et
al.~\cite{Friedman2007SpatialLife} as well. Additionally, results
showed that male dyads maintained larger interpersonal distances and
less mutual gaze than female dyads. They also reported that male dyads
were significantly less likely to maintain mutual gaze in indoor
locations which made sense due to the implication of higher levels of
intimacy associated with more mutual gaze and the constraints placed
on maintaining larger interpersonal distance in smaller
spaces~\cite{Yee2007TheEnvironments}. In a similar vein, Bonsch et
al.~\cite{Bonsch2018SocialEmotions} reported that participants chose
to maintain larger interpersonal distances to virtual agents perceived
to be angry in comparison to happy virtual agents.

``Personal space'' is clearly important in both physical and virtual
worlds. However, it's also clear that we don't have a complete picture
of proxemics and personal space when applied to virtual spaces. For
instance, unlike Bailenson et al.~\cite{Bailenson2001Equilibrium},
Llobera et al.~\cite{Llobera2010ProxemicsEnvironment} didn't detect
differences in participant physiological responses when they were
approached by cylinders. Takahashi et
al.~\cite{Takahashi2013PsychologicalEnvironment} proposed that cones
facing an observer were perceived as socially discomforting and
threatening. Nearly two decades ago, Bailenson et
al.~\cite{Bailenson2001Equilibrium,
  Bailenson2003InterpersonalEnvironments, Yee2007TheEnvironments}
suggested that proxemics could prove a valuable gauge for measuring
the behavioural realism of virtual agents/avatars. Recently,
McVeigh-Schultz et al.~\cite{McVeigh-Schultz2019ShapingVR} laid out
some design considerations, extracted from expert interviews with
creators of VR applications, which might shape social interactions in
shared virtual spaces including awareness of affordances of proxemics
and personal space management in order to tackle issues of
harassment. Positional data can now be easily surfaced in VR platforms
allowing us to use proxemics as part of a mixed-method evaluative
toolkit to, at the very least study, user experience in a bid to
design better spaces.


\subsection{Forming Groups and Social Activities} 

As with physical environments, the design of VEs
influences how people form groups and join activities. Inspired by
urban ethnography, Moore et al.\ describe environment design in terms
of accessibility, social density, activity resources, and
hosts~\cite{Moore2009FromSpaces}. For example, large open spaces can
be difficult to fill with enough people to achieve ``social
density.'' Virtual spaces that are not restricted by physical building
or space constraints often fail to create cosy places where social
contact comes easily.

Apart from influences of environmental designs, social cohesion and
social identity have long been recognised as two main factors that
impact on group
formation~\cite{Hogg1985InterpersonalFormation}. Social cohesion
traces group formation to processes of interpersonal
attraction~\cite{Hung2010EstimatingBehavior}, while the social
identity defines by considering identification, or self‐categorisation
to be the mechanism of group
formation~\cite{Hogg1985InterpersonalFormation}. Postmes et
al.~\cite{Postmes1999SocialGroup} found that, in computer-mediated
communication, where people do not necessarily have physical contacts,
groups, group identities and group norms can still be formed through
social interaction. Once the social identity is formed, anonymity
among group members in a computer-mediated world does not impede, but
mostly enhance group bonds and
performances~\cite{Postmes2001SocialBehavior}. For example, Ducheneaut
et al.~\cite{Ducheneaut2006Games} studied the group formation in a
virtual game world, and concluded that players usually remained
anonymous to each other and often formed groups because of their
complementary ``skills'', which could help them tackle difficulties
together in the game. Wessener and
Pfister~\cite{Wessner2001GroupLearning} argued that, for better group
performance, computer-mediated system should support three phases of
group formation, namely
\begin{enumerate*}[label=(\arabic*)] 
\item the initiation (e.g., giving prompts to form groups or assigning
  groups in advance);
\item identification (e.g., finding group members); and
\item negotiation (e.g., balancing the preferences of group members
  and the goals of the group).
\end{enumerate*}
The virtual environmental design and the pre-assigned groups of our
virtual workshop well considered these factors.

Moving from the physical world to VEs may change social activities:
creating new forms of interaction (e.g., use controllers to teleport
in VEs), and new social norms to keep the virtual world in
order~\cite{Moustafa2018AReality}. Ackerman et
al.~\cite{Ackerman2010SocialCode} studied social regulation in an
online game. They show specific social norms, defined to regulate the
game world interactions and technical interventions, are imposed in
game to automatically prevent or punish unwanted behaviours. Yee et
al.~\cite{Yee2007TheEnvironments} indicated that social interactions
in VEs are governed by the same social norms (i.e., social norms of
gender, interpersonal distance, and eye gaze) as social interactions
in the physical world.  Social VR has the potential to afford more
social interaction than video conferencing, such as the ability to
organically break off into small groups, or interacting with virtual
objects in the scene~\cite{Pidel2020CollaborationOverview}. Many
commercial platforms have implemented novel social mechanics
to stimulate social activities, such as designing a VE to simulate group discussion atmosphere, implementing
built-in tools to enable users to stay in VEs and focus on the social
tasks, or enabling users to use simple hand gestures to stop
harassment~\cite{McVeigh-Schultz2019ShapingVR}.

Avatar realism and co-presence experiences have been recognised as two
important factors for social VR activities. A recent study found that
a realistic VE created more presence, and a cartoon avatar created
stronger co-presence but was less trustworthy than a realistic
one~\cite{Jo2017EffectsSystems}. Steed and
Schroeder~\cite{Steed2015CollaborationEnvironments} indicated that
realistic avatar gaze may be important for one-on-one conversations in
VR, but avatar distinctiveness is probably more important than realism
when it comes to collaborating in large groups. Users often adjust
their behaviour to compensate for the lack of social cues. For
example, Roth et al.\ found that the absence of eye contact and facial
expressions shifts the user's attention to other the tone of voice,
and does not impede the task's
execution~\cite{Roth2016AvatarReality}. The compensations also include
exaggerating movements (e.g., flying up) when a user does not think
their collaborator can see them~\cite{Pidel2020CollaborationOverview}.

\section{Research Questions and Virtual Workshop
  Design}\label{sec:rese-quest-virt}
With the pandemic of 2020, numerous conferences and workshops were
cancelled or rapidly adapted for remote attendance.  Our workshop
design was planned as a hybrid with a virtual VR component and in person
component, but we pivoted
to a fully virtual workshop.  While redesigning our programme for virtual delivery, we explored three research questions:
\begin{enumerate*}[label=(RQ~\arabic*)] 
\item Are metaphors from proxemics in the phsycial world useful for analysing proxemics in a virtual world?
\item What is the relationship between groups, environment and
  activity in a virtual workshop?
\item What social cues or environmental artefacts help or hinder the
  virtual workshop experience?
\end{enumerate*}
Using data from our instrumented VE platform, observation notes, and
semi-structured interviews, we measured user proximity, behaviours,
and social experiences during the workshop.
Our focus was  to facilitate small group discussions, networking,
addressing open-ended problems, and building communities in focused
research areas.

\subsection{On-Boarding Sessions}
Prior to the workshop, we held on-boarding sessions to give participants a
chance to test their technical setup, ask questions about the event,
and familiarise themselves with Mozilla Hubs.  This was particularly important for those not
familiar with WASD keyboard movements.  WASD is a common design for
movement in video games~\cite{Moss:DarkCastle} which uses the inverted
T shape of those QWERTY keyboard keys for forward back left and right
movements which are coupled with a mouse for a secondary
action.  The on-boarding also introduced many participants to the VE
notion of flying, where a user can move into the sky and stay
stationary.  We also support participants during the opening the workshop with a warm up
session to address any last minute technical issues.  

\subsection{Workshop Programme}
The workshop officially started with a keynote and question session lasting forty
minutes. Following the keynote was a ten minute break. In the next session, each
participant gave a two minute pitch to introduce themselves in a
session lasting fifty minutes.  After a thirty minute break, the
participants, broken into groups, began discussions in small breakout
rooms accessed using portal linked from the main room. These discussions lasted one
hour with two breaks.  After the second break, the groups reconvened
in the main room to present their discussions and close the workshop
in a session lasting twenty minutes. The generous breaks throughout
the day allowed participants to rest from in browser or VR headset, and provided social interaction interludes akin to ``hallway
conversations''.

\subsection{Workshop Environments}
\begin{figure}
  \centering
  \subfloat[Aerial view shows the layout of the large meeting area,
  which covered an area of seventy by forty meters.]
  {\includegraphics[width=.45\textwidth]{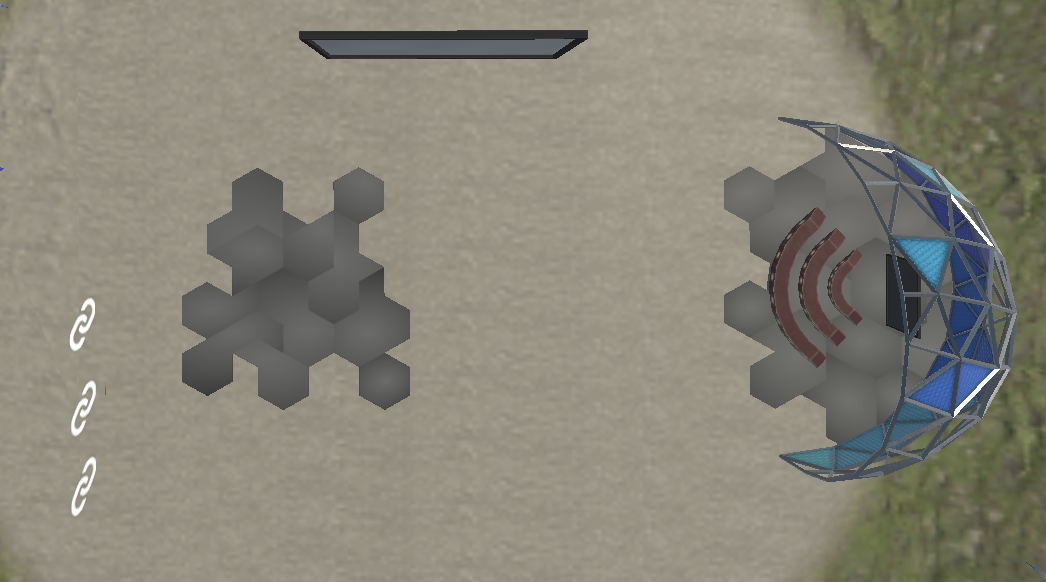}}
  \hspace{1pc}
  \subfloat[3D view shows the large screen and amphitheatre features
  available in the space.]
  {\includegraphics[width=.45\textwidth]{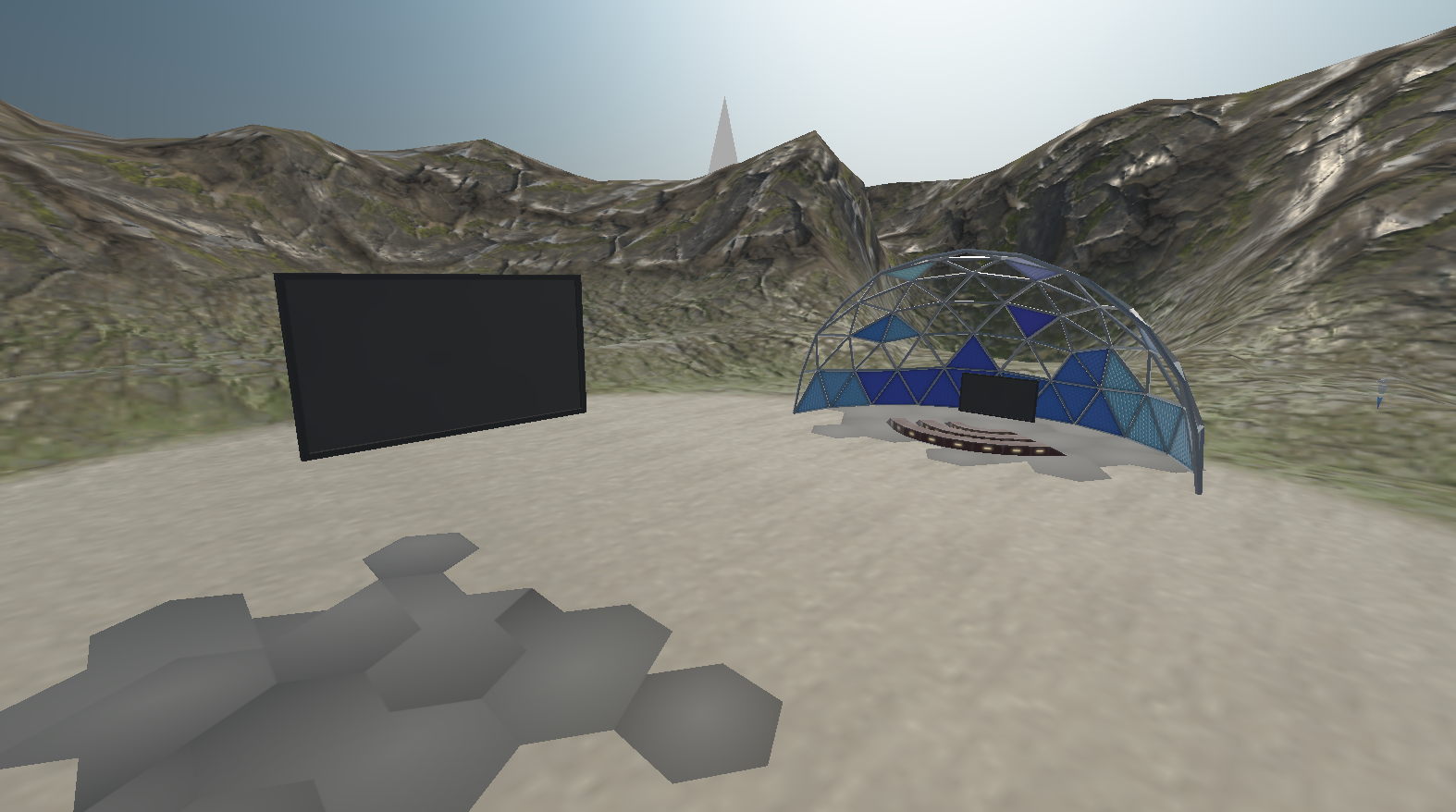}}
  \caption{\label{fig:outdoormeetup} \textit{Outdoor Meetup} was the
    larger meeting space where keynotes and informal breaks took
    place.}
  \Description{Left: Top down view of outdoor space.  A large virtual
    display runs along the top wall.  A dome amphitheatre is along the
    right wall.  A floor texture of hexagons if off-centre to the
    left.  Three portals to the breakout rooms are on the left wall.
    Right: A view of the large screen and the amphitheatre from the
    top of the bottom left corner of the room.}
\end{figure}

When designing a virtual world for the workshop, we opted for an outdoor lecture hall provided as a standard Mozilla Hubs space.  The VE asset
can be edited in a world builder from Mozilla called Spoke. This main
room, called \textit{Outdoor Meetup}, is a large open space with an
equivalent area of seventy by forty meters.  The standard scene has a large
display screen in the space, a smaller screen in a side
mini-amphitheatre, and several pre-populated tables. See
Figure~\ref{fig:outdoormeetup}.  We removed the tables as they only
block movement in the space and have no virtual affordances.  In the corners of the amphitheatre
space, we added some instructions for using Mozilla Hubs as well as some
Hawaiian themed items since the cancelled real-world event was to be in
Honolulu.  This room had some features disabled to keep
participants from cluttering the space with drawings or objects.

\begin{figure}
  \subfloat[Aerial View shows the layout of the twenty by thirty meter
  space.]
  {\includegraphics[height=.25\textwidth]{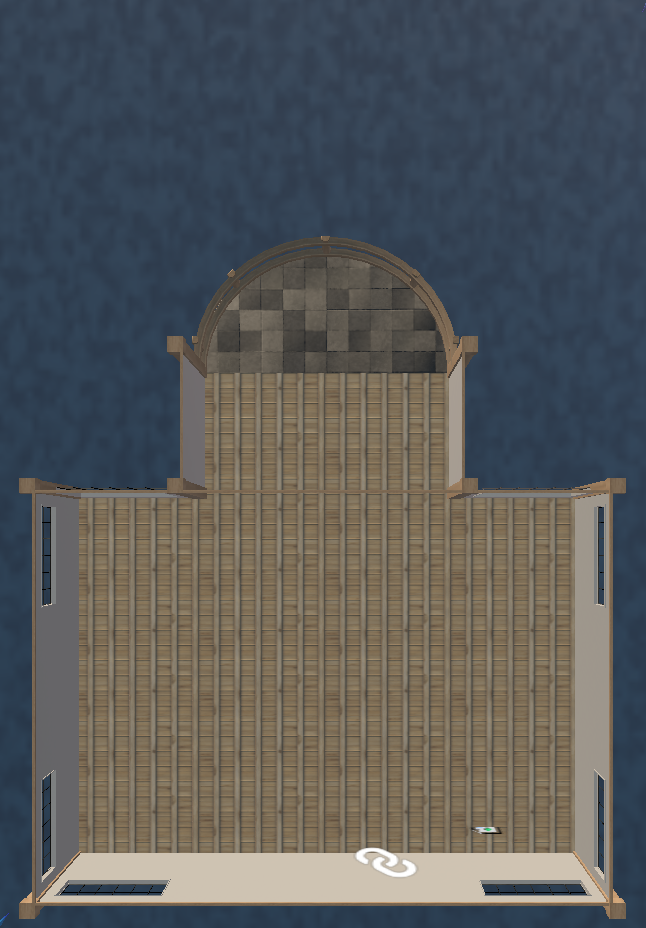}}
  \hspace{1pc}
  \subfloat[3D View shows the design of the \textit{Lake Office},
  including a small balcony overlooking a body of water]
  {\includegraphics[height=.25\textwidth]{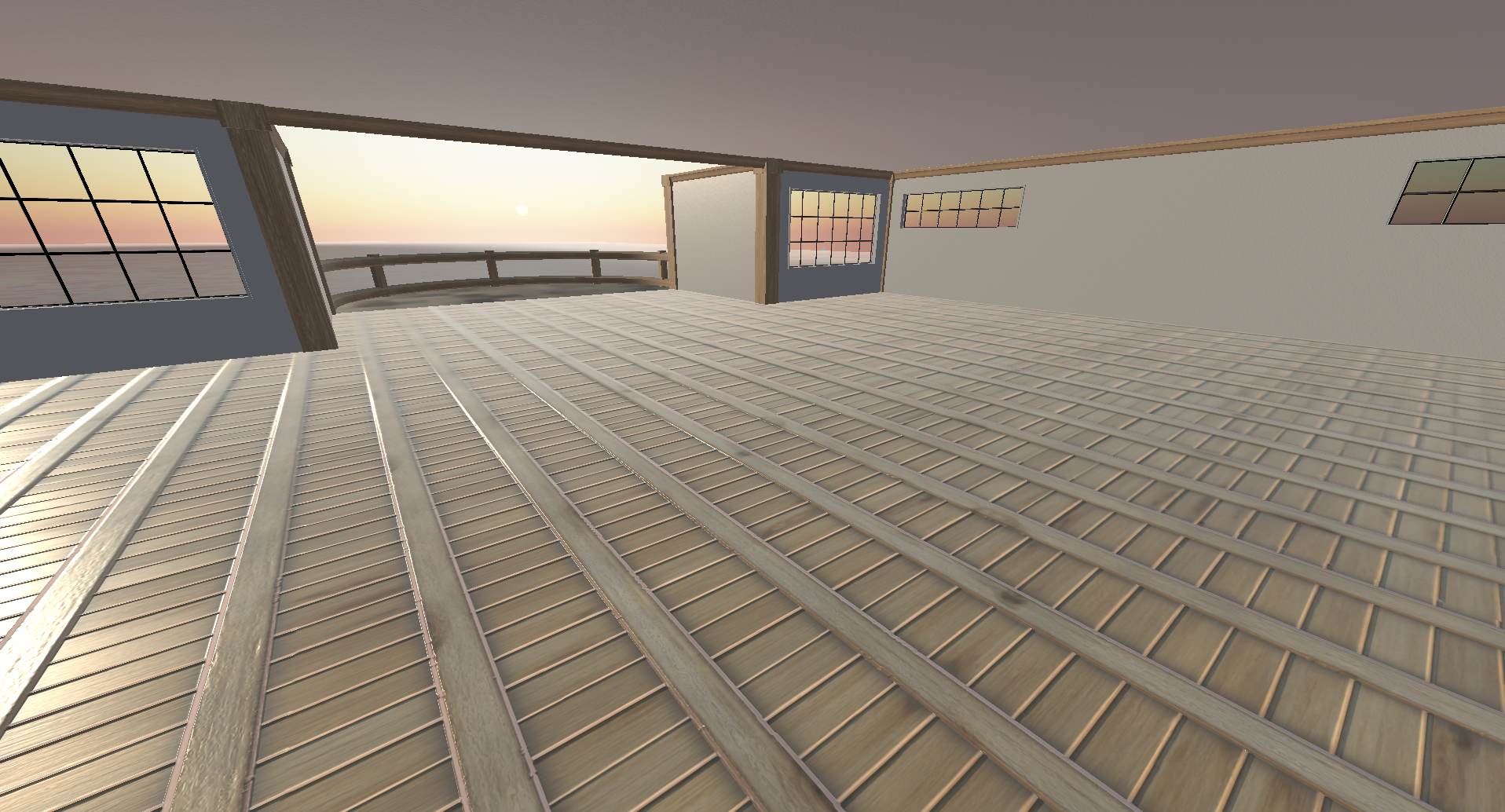}}
  \caption{\label{fig:lakeoffice} \textit{Lake Office} was a smaller
    meeting space, where small group discussions took place.  The
    three \textit{Lake Office} rooms have access portals,
    as shown in the south-west corner of the \textit{Outdoor Meetup}
    space in Figure~\ref{fig:outdoormeetup}.}
  \Description{Left: A top down view of the Lake Office.  The room is
    square with a semi-circle balcony extending from the top edge.
    The room is surrounded by shallow water.  Right: A 3D view of the
    room that looks out from the corner of the room towards the
    balcony.}
\end{figure}

There were three breakout rooms linked to the space with portals.
Each breakout room was identical based on another standard Mozilla Hubs scene
called ``Lake Office''. Each of these rooms had wall space for
screen-sharing and the room locks turned off so the breakouts sessions
could create objects, drawings, or other artefacts freely.  The Lake
Offices are equivalent to twenty by thirty meters (see 
figure~\ref{fig:lakeoffice}).  These breakout rooms had
a small balcony overlooking a shallow lake providing some indoor and
outdoor separation.

\subsection{Informed Consent}
To run this workshop while collecting data for our research, we asked
each workshop participant for informed consent before the workshop.
This consent included video recordings and photographs in the VE,
observational notes, and instrumented tracking.  This consent excluded
recording chat or logging audio (outside of the livestreamed keynote
on YouTube) as part of our privacy protections during the event.  We
explained steps we would take to anonymise the data, including removing
participant names that would appear within the VE\@.  Participants had
the option to opt out of the semi-structured interviews.

\section{Instrumentation, Data, and Interviews}
Mozilla Hubs is an experimental, VR-friendly platform from Mozilla
Mixed Reality. Mozilla Hubs works with immersive head mounted displays
like Oculus Quest and traditional web browsers on a computer or mobile
device like an iPhone. To collect data from our workshop we modified
the Mozilla Hubs Client to track and log positional data, recorded
observations during the workshop, and conducted semi-structured
interviews after the workshop. The code described in this section for
Mozilla Hubs Cloud data collection, the collected dataset, and scripts
for analysis is openly available (see Appendix~\ref{app:dataset}).

\subsection{Instrumented Virtual Environment}\label{sec:instrument}
The Mozilla Hubs platform itself is entirely open source which allows
people to stand up their own server instance with a custom client
(called Hubs Cloud) on Amazon Web Services or Digital Ocean. We used a
Hubs Cloud instance to run the workshop with instrumentation to
facilitate the data collection.  There is a Janus WebRTC Server for
keeping track of all the messages, objects, and users in each room in
Mozilla Hubs.  To track where every user is, one could write code to
constantly query the Janus server.  However, this would create added
load on an already stressed server that reaches its maximum capacity
at thirty users. It is possible to modify and stand up a custom Janus
server but that is generally not advised in the documentation.  For
our solution, we turned to client side logging.

Mozilla Hubs uses a two core web based frameworks, A-Frame and
three.js, on the client.  A-Frame is a web and HTML framework for
virtual reality (WebVR); three.js is a cross-browser JavaScript
library for creating and displaying animated 3D. These two provide the
rendering engine that runs Hubs.  More importantly, one can subscribe
to \texttt{tick} events in the browser and on each tick log the needed
data:
\begin{enumerate*}[label=(\roman*)] 
\item A unique identifier of the player
\item Current UTC timestamp
\item Have they entered in the room or are in the lobby?
\item Where are they in the room (position vector as x, y, z)?
\item Where are they looking/facing in the room (direction vector, orientation quaternion)?
\item What is the client's current rendered Frames-per-second?
\item Are they muted?
\item\label{sec:instr-virt-envir:talking} Are they talking (and if
  so, how loud on their input mic)?
\item\label{sec:instr-virt-envir:damp} Do they have the spatial audio
  dampened?
\end{enumerate*} The overall framework allows one to add other
listeners for events like use of the laser pointer or drawing pen.

Next, a HTTP server collected data from each client's POST data.
However, each client logs data on every tick; this collection would
add up to tens of tick POSTs per second. This would cause socket
failures on the client and server.  Here an open web-socket would
allow data to flow freely. Instead, we opted for batch collecting
ticks on the client then doing a bulk POST operation.  We empirically
tested the server and set a threshold of 4000 ticks per post.  The
collection server was a simple nginx HTTPd server with a python uWSGI
application which logged JSON data via concatenation to disk. 

\subsection{Proxemic Dataset}
Overall we collected 17,779 user input events (which includes the
pose-position vectors) and 13,928 user info events (which track device
usage). For this event, the logger captured positions every 1000 frames rendered. As the participants were on devices of different
capabilities, their effective frame-rates varied
($\mu=43.67, \sigma=15.3$) resulting in some participants having
logged more frames than others.  To address this, the pose-position
data was resampled to ten frames per minute for standardisation.
The resampled dataset had 37,457 user position logs organised into
6,619 frames from 26 users across four virtual spaces.  Although the workshop had 27 registered attendees, one participant experienced technical trouble and could not attend the event.  

The resampling process organised logs into frames, which are mapped to timestamps in UTC\@.  For each frame, we calculated pairwise distance (euclidean in 3D space) and angle (based on position
coordinates and direction vector) between each person.  This is
represented as two dictionaries of matrices, indexed by frame. 

\subsection{Observations}
The workshop lasted 4.5 hours on April 29, 2020, which started at
16:10 (GMT+2) with co-organiser's introduction, and ended at 20:40
(GMT+2) with a virtual group photo. All the planned workshop
activities, including the keynote speech, individual pitches, social
breaks, group discussions and final presentations went smoothly. Some
technical problems did exist, especially during the first half an
hour, when a few participants were frequently reporting frozen frames
and interrupting audio. However, these problems were solvable by
refreshing and rejoin. Interestingly, even during
social breaks, many users (14--18 users) chose to stay in the virtual
world and network with each other. No participants reported technical
problems during the group discussion in the breakout rooms. The
participants in breakout room A spent their social breaks by jumping
together into the virtual ocean.

\subsection{Semi-structured Interviews}\label{sec:semi-struct-interv}
After the workshop, we conducted semi-structured interviews with nine
participants (P1-P9, 6 females and 3 males). We aimed to collect participant feedback on overall experience
such as quality of interaction/communication, their evaluation of Hubs
and their thoughts of how the workshop compared to others they might
have attended in the real world. Each interview was forty-five minutes
long. In order to guide the conversation, we organised each interview
into three parts.  First was an opening question to encourage
participants to talk about their overall experiences with prompts to
remind them of specific sessions in the workshop to talk about for
instance `the keynote presentation' session or `the breakout session'.
Next, there was a discussion of the participants experience in the
virtual workshop (in comparison to physical workshops if relevant),
what delighted them about it, their frustrations about what did not
work, the ease with which they partook in conversations, their ability
to express themselves and their ability to form impressions of others.
Finally, there were some user experience style questions asking for
feedback about Hubs, in comparison to other video conference
platforms, w.r.t audio quality, visual representation of the space,
creation of avatar, usage of chat/emojis, and water cooler moments (if
any).

\section{Results}
Our results bring together analysis of the 26 users logged in the
proxemic dataset (3 out of 26 users used HMDs), field
observations, and 9 individuals given post-workshop interviews.  Of
the interview participants, eight attended the full workshop.  P4 had
significant technical difficulties during the workshop, but attended
the workshop on-boarding session. P6 and P7 used an HMD (i.e., Oculus
Quest) for the workshop, the other six participants all used web
browsers (i.e., Chrome and Firefox). 



\begin{figure}
    \centering
    \includegraphics[width=0.45\textwidth]{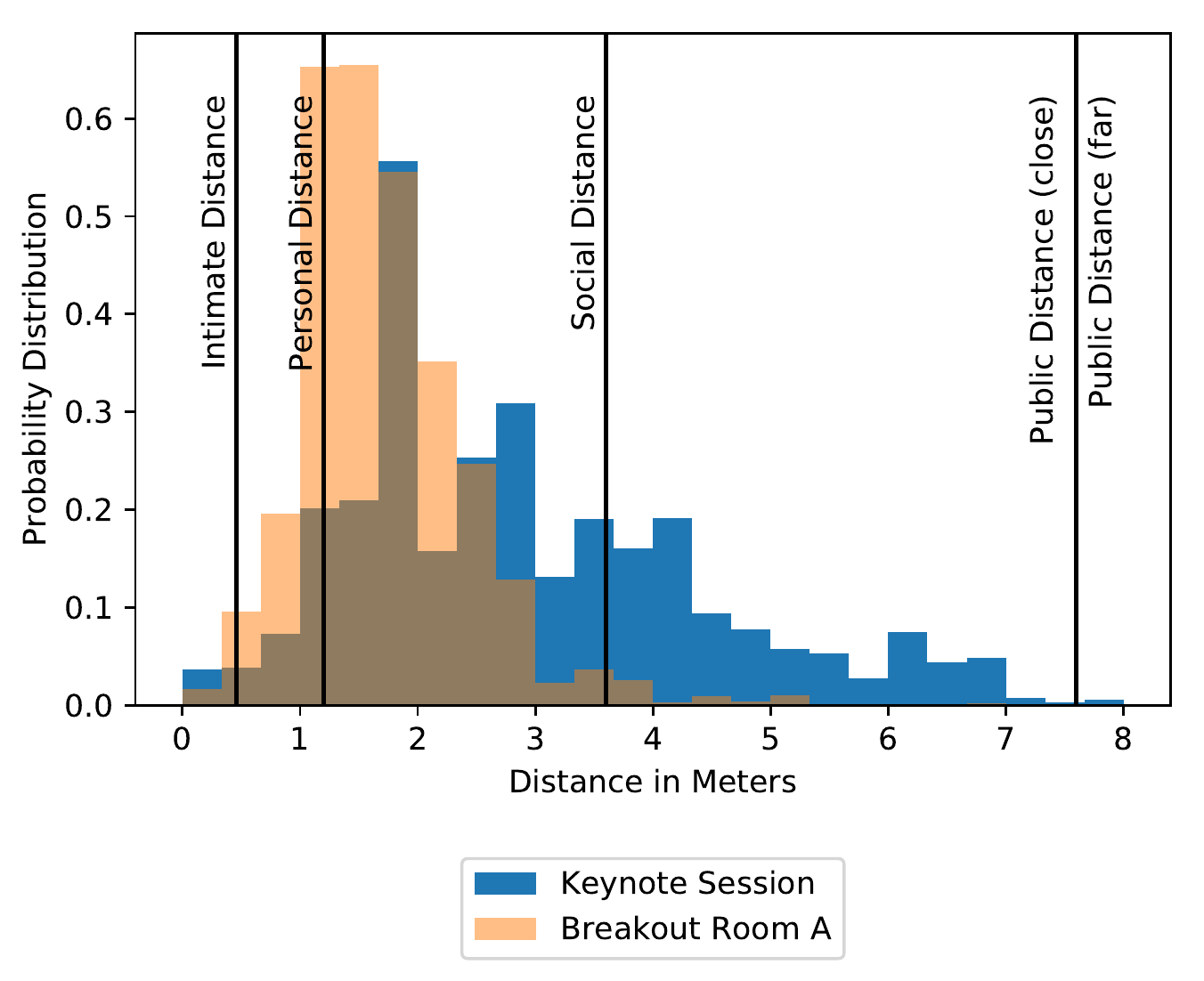}
    \caption{\label{fig:closet_person} This histogram shows the
      probability distribution of the nearest participant (euclidean
      distance in meters) in the Keynote Session compared to Breakout
      Room A. The proxemic zones~\cite{Hall1969ThePrivate} and
      labelled across the X axis.  In the small group discussions in
      Breakout Room A, participants stood closer. There was a greater
      probability of standing with others in the personal and social
      distances as compared to the Keynote session.  Collisions in the
      intimate distance occurred infrequently in both sessions and
      participants adjusted their social spacing when intimate
      distances occurred.}
    \Description{A bar chart with two data series; next nearest person
      during the keynote and next nearest person during the breakout
      session in Room A.
|Keynote | Room A |
|--- | --- |
|0.02851382 | 0.01699235|
|0.03542627 | 0.09260833|
|0.07430876 | 0.19031436|
|0.11751152 | 0.64655905|
|0.19009217 | 0.64740867|
|0.59360599 | 0.55480034|
|0.15552995 | 0.35683942|
|0.18577189 | 0.24468989|
|0.36895161 | 0.13508921|
|0.16417051 | 0.02209006|
|0.16330645 | 0.04078165|
|0.18663594 | 0.02548853|
|0.22292627 | 0.00254885|
|0.0812212  | 0.00764656|
|0.06480415 | 0.00339847|
|0.05529954 | 0.01019541|
|0.09245392 | 0.        |
|0.02073733 | 0.        |
|0.09158986 | 0.00084962|
|0.04752304 | 0.        |
|0.04406682 | 0.00169924|
|0.00604839 | 0.        |
|0.00432028 | 0.        |
|0.00518433 | 0.        |}
\end{figure}

\subsection{Social Cues and Proximity in Virtual Space}
Using Hall's proxemic zones~\cite{Hall1969ThePrivate}, we analysed
participants' proximity and personal space in the VE\@.  Our goal was
to explore whether these zones were a useful construct for
understanding how interaction unfolded during different workshop
activities.  Although previous research has shown the utility of
measuring proxemic zones in VR, our dataset provides a continuous
tracking at a high resolution to enable new kinds of
analysis. 

Figure~\ref{fig:closet_person} shows the probability distribution of
collisions with the next nearest participant in the ``intimate zone''
(less than 0.5 meters) during the keynote session and breakout
discussion in Room A.  Mozilla Hubs does not prevent collisions or
enforce ``personal space'' in software, but collisions in this space
are still relatively rare.  Such collisions can easily occur on
``spawn'' points when users enter the room at the same coordinates,
but participants quickly moved away from others.  Although
participants discussed the importance of personal space, they also
discussed some challenges. For example, some participants pointed out
that it was not easy to perceive the positions and distance,
especially to sense the things behind their avatar (P2, P3,P8).

We observed different behaviours across the proxemic zones during
different workshop activities. Figure~\ref{fig:closet_person} shows
the probability distribution of the next nearest participant during
the Keynote Session and Breakout Room A.  In Breakout Room A,
participants formed cohesive small groups, where the next nearest
participant was within the \textit{personal} or \textit{social}
distances.  During the Keynote session, the participants formed less
cohesive groups, where the next nearest participant extended further
away into the \textit{public} zone.  These insights demonstrate that
impacts from activities or environments could be observed in the
proxemic data.

Mozilla Hubs provides spatial and audio cues based on each person's
position, field of view, hand pose (if using an HMD), and
microphone. Participants described the challenges of interpreting
others' intentions or starting conversations without additional cues
like facial expressions and eye contact. P9 stated that \textit{``I
  didn't know when to cut into people's talk, to express my
  opinion.''}. Similarly, P3 stated that \textit{``I wasn't sure
  whether people were paying attention. I would describe it as the
  biggest challenge in social VR where you don't have facial cues like
  eye direction.''} Audio cues were the most problematic during the
workshop. All the participants mentioned that they couldn't have
private chat or were uncertain how far away others could hear
them. Audio often overlapped, and participants didn't want to
interrupt others. If they turned on the spatial audio and moved away
from others to chat, it could be hard to hear the workshop organisers
and rejoin the main group. As P4 said, \textit{``You just move from
  the main group and then you could have a private conversation. I
  know it is possible, but I didn't do it. In this [the social VR]
  setup, I was really focused on staying close to the group, not
  missing anything important.''}

HMDs users benefited from additional social signals with tracked head
and hand movements.  Participants noticed how much more expressive
participants in HMDS could be (P1--P3, P6--P8).  P7, who wore an HMD,
stated that \textit{``wearing an HMD with moving hands made us more
  expressive. We seemed more important, because we can give more
  information and attract more attention using hand gestures.''}
  Three participants (P2, P6, P7) wearing HMDs discussed how
they perceived personal space and felt they could better pick up on
social cues. P2 commented, \textit{``I was not entirely aware [of the
  interpersonal space] because I couldn't easily turn my head to look
  behind me. So, I wouldn't have known if my avatar was standing right
  in front of somebody's eyesight. But I was conscious and always
  trying not to invade anyone's personal space.''} All the
participants mentioned that, it was much easier to sense the social
cues and to have conversations in the breakout rooms. There was less
audio overlapping, fewer people and smaller space where people cannot
\textit{fly} high. Group members were standing in a circle, which was
easy to immediately identify the speaker. The circle felt like a
realistic spacial relationship as in a real workshop. As P6 said,
\textit{``In the small room, it was more comfortable to talk because
  we were standing in a small circle. You could just tell people's
  intention from the small movements of the avatars. It was
  comfortable to interact with people.''}

\subsection{Groups and Social Interactions}

\begin{figure}
    \centering
    \includegraphics[width=0.45\textwidth]{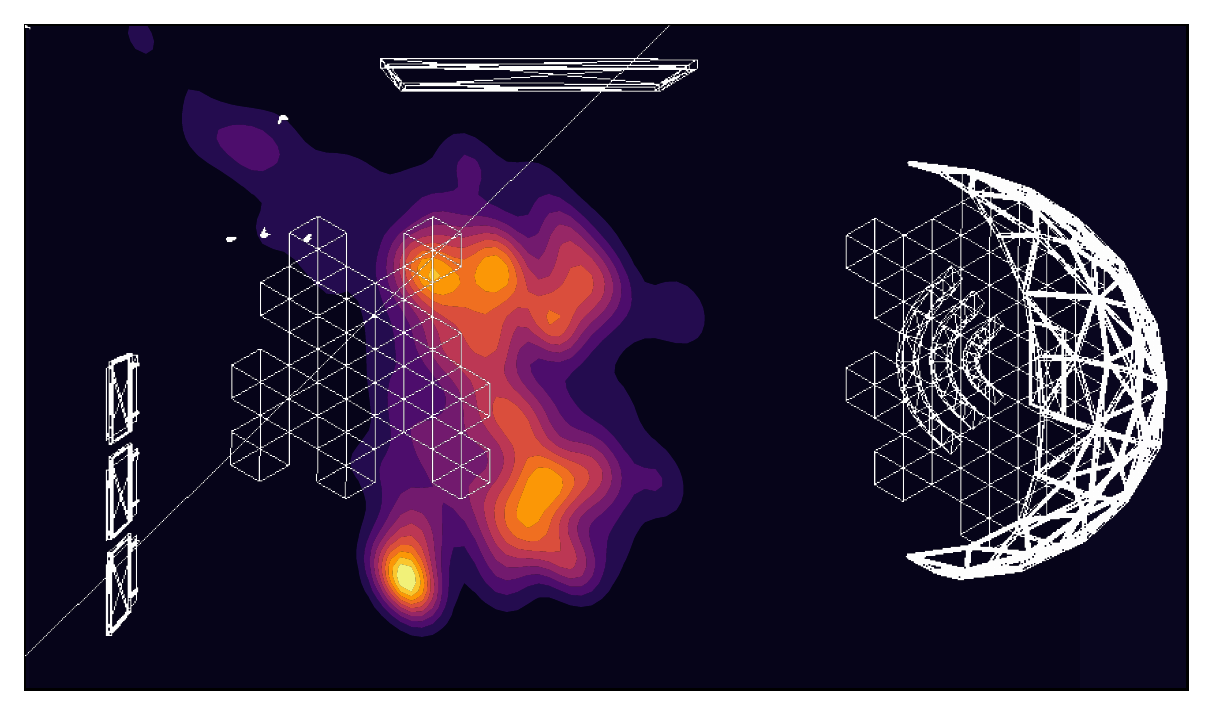}
    \caption{\label{fig:pos_overview} Position data overview for the
      duration of the workshop in the \textit{Outdoor Meetup} space.
      Activity took place most often in the centre of the room, but
      the floor decals in the space were an unexpected deterrent.
      Participants also avoided the large amphitheatre and no
      activities made specific use of this feature.}
    \Description{Top down view of the Outdoor Meetup space with a
      heatmap.  Hot spots on the heatmap occur in the centre of the
      room and avoid the hexagonal grid pattern.}
\end{figure}

Figure~\ref{fig:pos_overview} shows all the positional data for the
duration of the workshop in the \textit{Outdoor Meetup} space.  This
was a large open space with activity focused in the centre of the
area.  The floor decals, visible as the hexagonal grid in
Figure~\ref{fig:pos_overview}, appear to have discouraged participants
from standing in those areas.  Pedestrian traffic is incredibly
sensitive to texture and appearance in physical
settings~\cite{Whyte1982TheSpaces}, and VEs create similar
experiences.  The amphitheatre in the space was also poorly utilised,
although we did not organise any activities that made use of it.
Given the limited interactivity and resource provided by the static
amphitheatre, this is in line with previous
research~\cite{Moore2009FromSpaces}.

\begin{figure}
  \centering
  \includegraphics[width=0.45\textwidth]{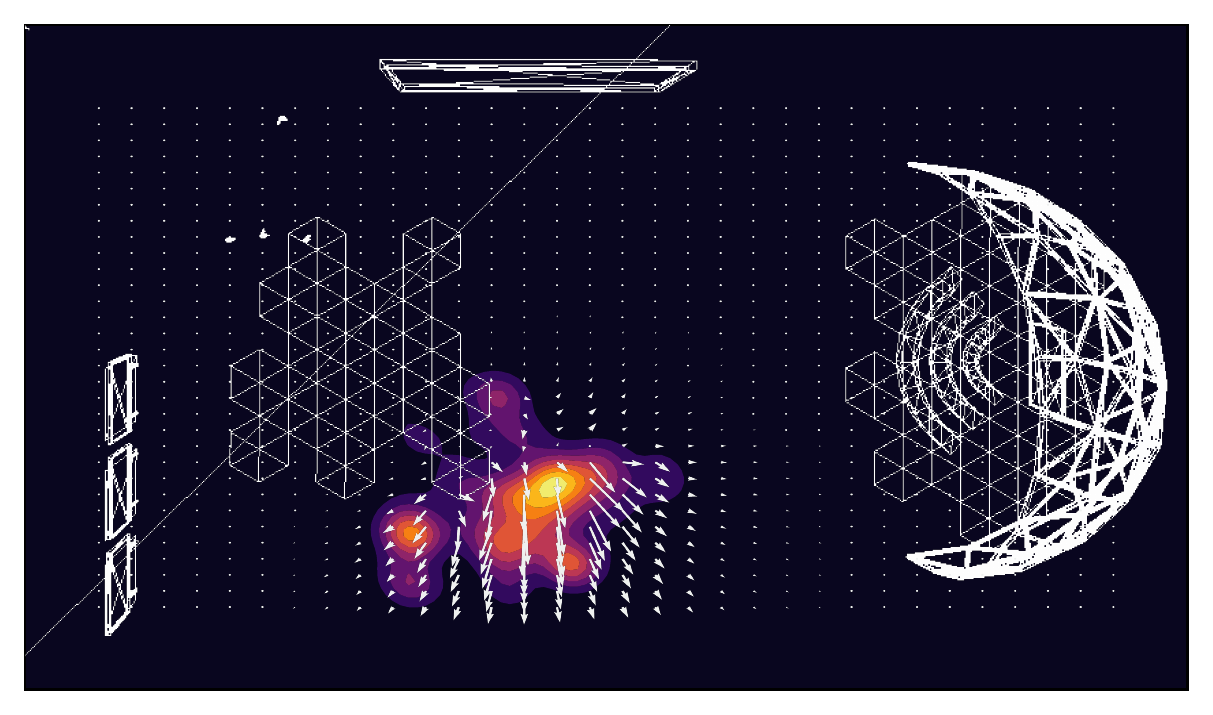}
  \caption{\label{fig:keynote_quiver}Heatmap of the \textit{Outdoor
      Meetup} space during the Keynote Session overlaid with a quiver
    plot visualisation participants' field of view.}
  \Description{Top down view of the Outdoor Meetup space with a
    heatmap and quiver plot.  Heatmaps shows hotspots in the centre
    along the bottom wall.  The quiver plot has arrows which radiate
    out from the hotspot towards the wall where the presenter stood.}
\end{figure}

\begin{figure}
  \centering
  \includegraphics[width=0.45\textwidth]{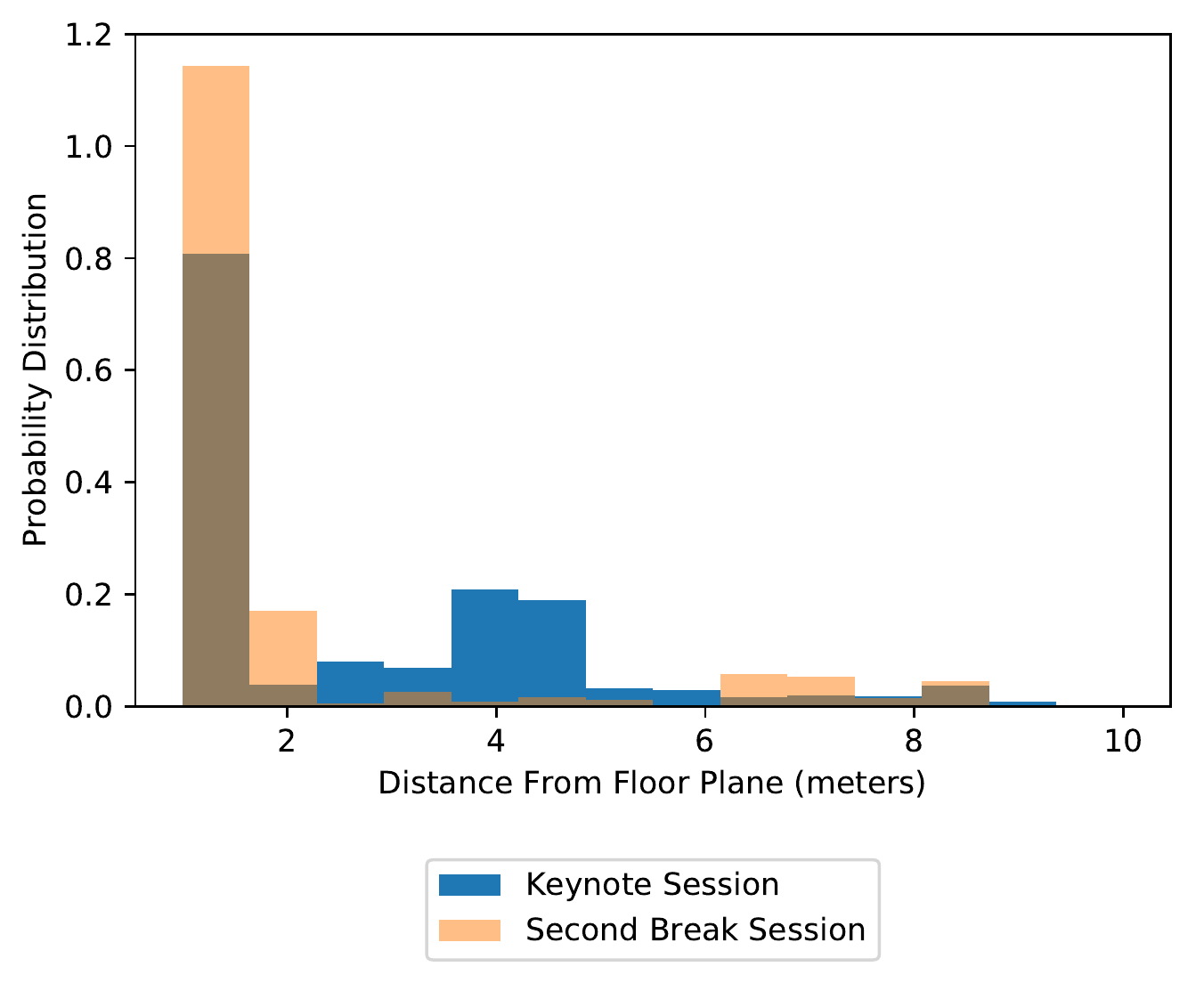}
  \caption{\label{fig:offthefloor} Histogram of z-axis in
    \textit{Outdoor Meetup} during the keynote and the second break.
    With Fly-mode enabled, participants could use the vertical space
    to gain a better view, but during breaks participants were more
    likely to constrain themselves to the floor plane.}
  \Description{A bar chart with two series; the Y positions during the
    key and during the second break session.
   | Keynote | Second Break Session|
   | --- | --- |
   |0.82414995 | 1.15066523|
   |0.02769144 | 0.17619561|
   |0.06417381 | 0.00575333|
   |0.07604157 | 0.0273283 |
   |0.20702647 | 0.00862999|
   |0.20219145 | 0.01654081|
   |0.03560328 | 0.01150665|
   |0.02813098 | 0.00287666|
   |0.01846096 | 0.06040992|
   |0.02197733 | 0.0546566 |
   |0.0131864  | 0.01150665|
   |0.02857053 | 0.0294858 |
   |0.00835139 | 0.        |
   |0.         | 0.        |}
\end{figure}

During the keynote session, the speaker displayed their slides in the
centre of the southern wall in \textit{Outdoor Meetup}, opposite from
the large display embedded in the environment.
Figure~\ref{fig:keynote_quiver} shows a heatmap of the participants'
positions and a quiver plot of their field of view during the keynote
presentation.  During the keynote, participants kept the speaker
within their field of view 54\% of the time and maintained this as a
shared point of attention. With plenty of room to spread out in the
seventy by forty meter space, participants still occupied a relatively
small area, 20$\times$15 meters of the 70$\times$40 room with sixteen
people. Participants used the z-axis ``fly mode'' to simulate graded
stadium seating while still maintaining personal space, as shown in
Figure~\ref{fig:offthefloor} where we measure proximity using 3
dimensions.
The discussion breakout rooms were smaller with low ceilings, as seen
in Figure~\ref{fig:lakeoffice}.  This resulted in closer distances
amongst the participants (Figure~\ref{fig:closet_person}) and limited
use of flying within the space (Figure~\ref{fig:offthefloor}). In
Mozilla Hubs, participants could fly through the ceiling and still
fully participate in discussion, but this kind of behaviour was not
observed. Intimate space collisions were also infrequent in the
breakout rooms, as shown in Figure~\ref{fig:closet_person}.

\begin{figure}
  \centering
  \subfloat[First Break]{\includegraphics[width=.45\textwidth]{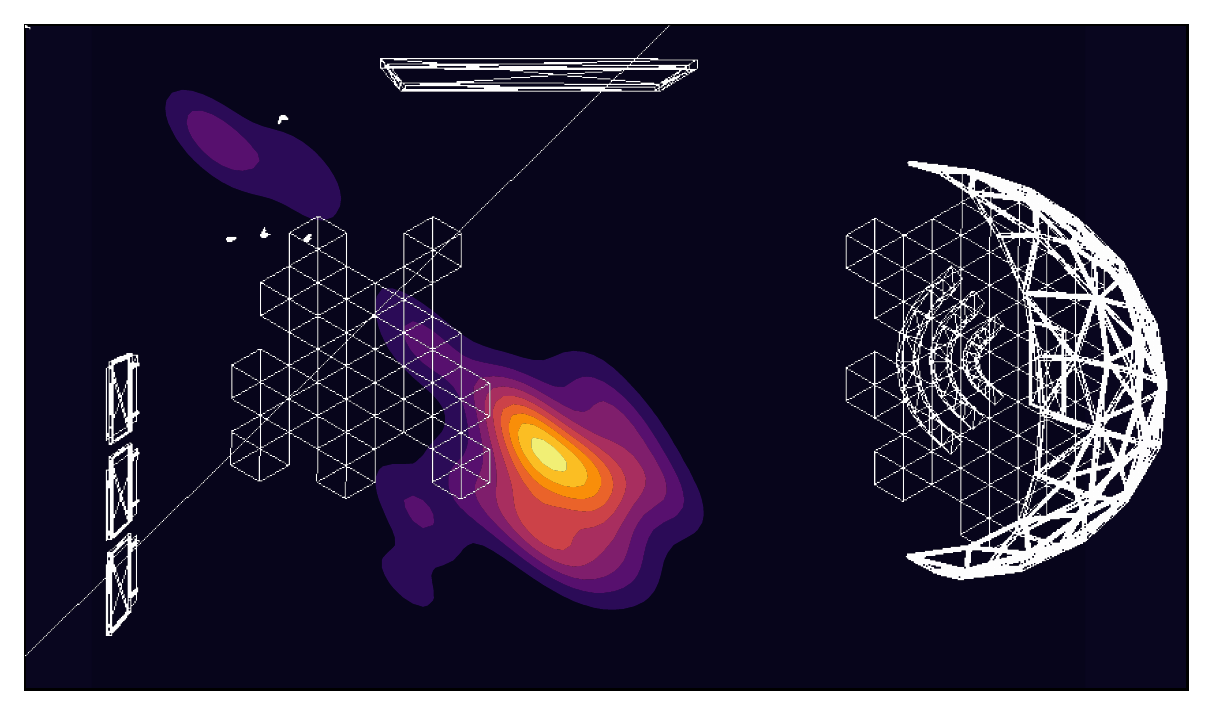}}
  \hspace{1pc}
  \subfloat[Second Break]{\includegraphics[width=.45\textwidth]{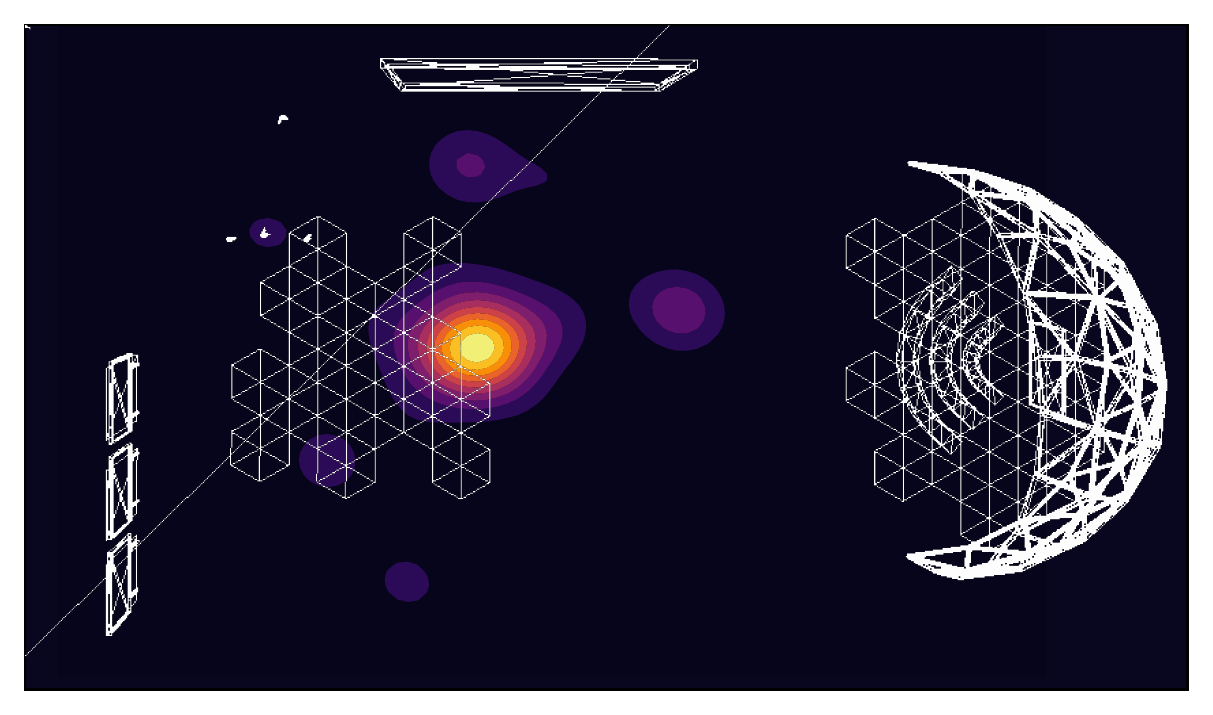}}
  \caption{\label{fig:breaks_map} Heatmaps of the \textit{Outdoor
      Meetup} space during scheduled breaks show where participants
    formed groups. During the first break, there was only one large
    group.  During the second group, small groups surrounded a large
    group.}
  \Description{Left: Top down view of Outdoor meetup with a heatmap.
    There is a large central hotspot with one smaller hotspot in the
    top left.  Right: There is a medium central hotspot surrounded by
    5 small hotspots.}
\end{figure}

The workshop included four breaks to reduce fatigue and give
participants time away from their screens. However, many participants
stayed in the virtual space for informal conversations and networking.
Figure~\ref{fig:breaks_map} shows how participants formed groups
during the first and second break sessions in the \textit{Outdoor
  Meetup} space.
At the first break, participants formed a larger group that didn't
have clear boundaries. During the second break, there was a large
group surrounded by smaller satellite groups that became distinct
points.  While on break, participants were also more likely to remain
on the floor plane, as shown in Figure~\ref{fig:offthefloor}.  As
participants became accustomed to the controls, using the virtual
space, and sounds, they could form meaningful smaller groups.  For
example, participants figured out ways to attract attention or show
that they were paying attention, such as \textit{flying} to the
visible positions to gain attention (P1--P9) or turning their avatar to
face a speaker to show attention (P1--P9).

\begin{figure}
  \centering
  \includegraphics[width=0.45\textwidth]{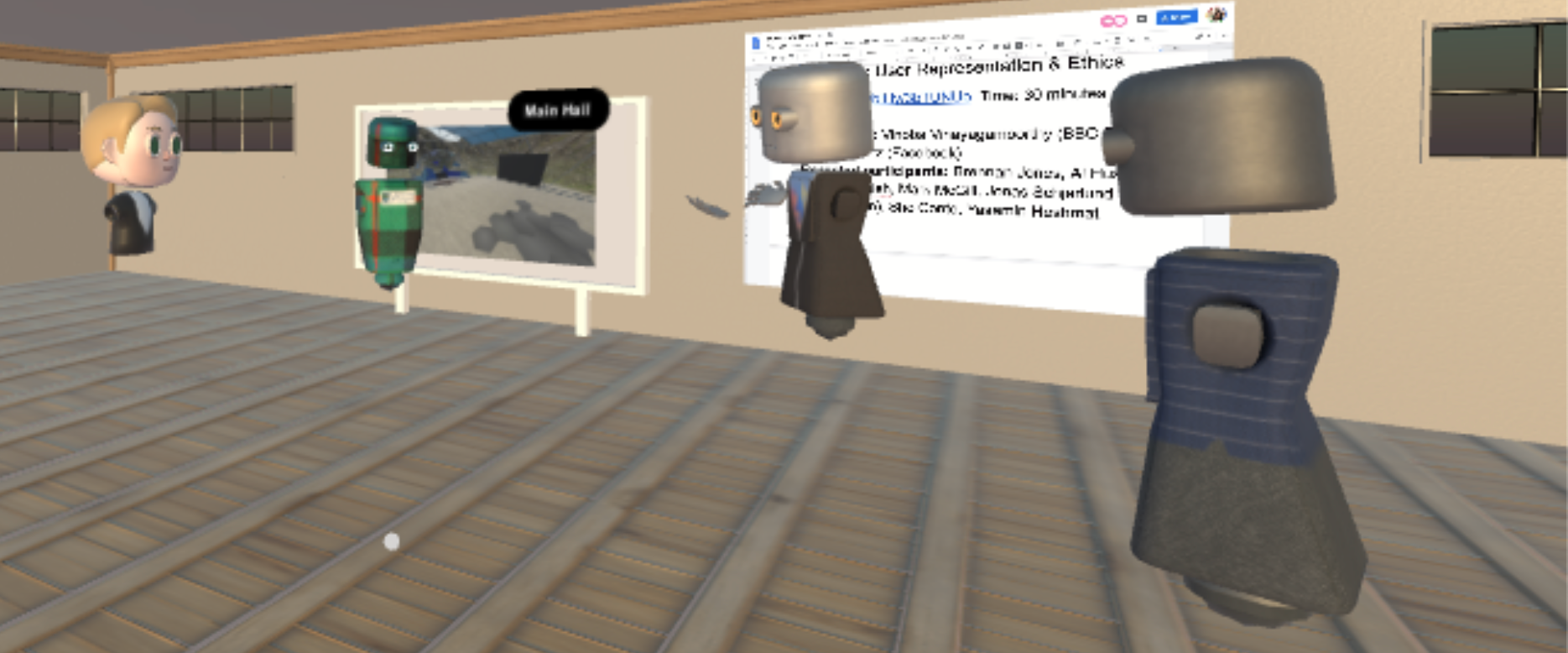}
  \caption{\label{fig:teaserppl}Workshop participants during a small
    group breakout.  The avatar second from the right has hand
    tracking enabled.  Avatars without hand tracking have no hands
    rendered in the virtual environment.}
  \Description{Four avatars in Mozilla Hubs collaborating around a
    document whiteboard. The avatar closest to the whiteboard has hand
    tracking enabled.}
\end{figure}

The participants discussed how their virtual workshop experience
compared with face-to-face social interactions. When networking,
participants found that the avatar faces the VE made it difficult to
associate the avatar with the actual person. Participants felt there
was additional effort to keep in contact with the people that they met
during the workshop (P3, P5, P6), and they had fewer opportunities to
encounter people by chance and start a conversation (P1, P2, P7). As
P3 said, \textit{``If we meet in person, and later we meet again, I
  would feel comfortable talking them, versus if we meet in avatar
  forms, next time when we actually see each other in person, we would
  still feel a need to reintroduce ourselves.''}  Although we did not
specifically address avatar capability or customisation in this
research, participants discussed their experience with avatars during
the workshop. The avatars in Mozilla Hubs are typically cartoonish or
robotic representations, as shown in Figure~\ref{fig:teaserppl}. Most
participants spent time selecting their avatar from Mozilla Hubs's
defaults or customising using a third party
tool\footnote{\href{https://rhiannanberry.github.io/Avatar-Customizer/}{Avatar-Customizer}
  used by the IEEEVR 2020 VR Conference, retrieved on Aug. 28,
  2020.}. The avatars are not photo-realistic, but the customisation
allowed participants to represent themselves with a chosen
appearance. Some participants pointed out that they prefer their
avatars to be realistic with expressive body language. They felt this
would help them network and remember others more easily (P5, P7,
P8). Other participants felt relaxed about their virtual appearance,
didn't worry about others' judgement, and felt de-indentified but
well-represented by their avatars (P3--P6, P9).  P4 stated that
\textit{``somehow having the characteristics related to myself, but
  not myself.''} The avatar worked as an ``equalizer (P6)'', which
enabled participants to be less anxious to approach senior people.

\subsection{Play and Finding Familiarity}
Participants used the time between activities to network and play
creatively in the virtual space, for example jumping into the virtual
ocean (P2, P6, P7-P9) and taking virtual selfies (P2, P8, P9). These
playful activities became an effective icebreaker, for example P9
stated that \textit{``before we jumped into the virtual ocean,
  everyone was talking and behaving professionally, a bit stiff. You
  hardly connect with people. But this [jumping into the virtual
  ocean] enables you to connect with everyone.''} Participants also
mentioned that the selfie feature was useful and enabled them to keep
virtual memories, feel like they had been in a new place, and bond
with the other participants (P2, P3, P5, P6, P8). P2 stated that
\textit{``taking selfies in social VR is definitely more interesting
  than a screenshot in video conferencing. You feel like you are
  actually in a new place, which is a big deal when you are in
  quarantine for a while (P2).''}
Mozilla Hubs was new for most of the participants.  This added
difficulties for them in navigating the environments and utilising all
the available features.  The novelty of the VE could be distracting,
for example P7 stated that \textit{``the theme of the workshop is
  social VR\@. So, the Hubs itself became a conversation trigger,
  which is good. If it is a workshop about another topic, I am not
  sure the social VR platform would be more distracting than
  helpful.''}  Participants also mentioned difficulties such as saving
or removing objects and selfies (P4, P7); feeling unnatural to turn
head to see things (P3, P7); and difficulties in navigating to exact
positions (P1, P3, P4).

The on-boarding session before the workshop was helpful for the
participants to learn basic controls and social mechanisms in social
VR (P1, P3, P8). As P3 mentioned, \textit{``I was glad, throughout the
  training session, I had an opportunity to practice different
  controls. I wasn't sure how to figure out which people was taking,
  and later on, I learned this kind of social cues.''} All
participants mentioned that, when first entering the space, they moved
around and explored the virtual space. They usually stood still when
the speakers started to talk. \textit{``When I entered, it was empty,
  so I moved around to just explore the virtual environment. When
  there were things happening, I mostly just stood at one point to
  have a good view (P6).'' }
Novelty also meant that participants weren't always sure what they
were ``allowed'' to do or how they should behave. For example, P1
stated that \textit{`the whole [social VR] experience at the beginning
  felt almost like my first conference experience. I was not really
  sure what I am doing, where I am, how to behave, how the space
  works, how loud my audio is and so on. The training session prepared
  us technically, but not psychologically (P1).''}  Although the VE
resembled face-to-face interactions in many ways, the social cues
where not always enough to establish social norms.  For example,
physical proximity and personal space were easily managed using visual
cues.  Speech, however, was harder to control as the sound falloff did
not always behave as it would in a face-to-face interaction.

\subsection{Limitations}
The workshop was a live event on the topic of Social VR\@; many of the
participants were, of course, interested in the domain.  However, 7
had no VR experience, 7 reported using VR once or twice briefly, 3
said they used VR several times, and 1 person stated they were a VR
expert.  As everyone was curious about VR and there to learn more,
this shapes some of our qualitative findings.  There were some
technology failures and unforeseen limitations.  We completed an on-boarding session to help workshop participants familiarise themselves
with Mozilla Hubs and test their hardware before the workshop.  Many
participants still experienced challenges, such as frequently dropping
out (P1--P4) and audio issues (P5, P6, P8) likely attributed to
work-from-home owners and smaller home Internet bandwidth connections.
At the time of our study, Mozilla Hubs recommended a maximum of
twenty-five participants per room for optimum
performance\footnote{\href{https://hubs.mozilla.com/docs/hubs-faq.html}{What
    is the capacity of a Hubs room?}, accessed September 2020.}.  This
is variable based on client setup, but some of the participants
experienced technical issues even with fifteen participants in the
room. We also chose not to complete controlled comparisons, for
example comparing equal numbers of browser and HMD users, during this
live event.  Our results are based on the unconstrained behaviours of
the workshop participants, but future work could use our instrumented
approach to complete more controlled experiments based on high
precision proxemic logs. It would also be interesting to add audio
signals to our analysis of proxemic logs to investigate the variations
in spatial behaviour in response to aural differences.  Although we
subsequently added microphone volume tracking to our instrumentation
code, this was not available during the workshop.

\section{Discussion}
Open source VEs allow for unique instrumentation opportunities that
can help advance research on social interaction in VEs.  This paper
reports an initial step in that direction, which we hope will support
future work and enable new kinds of evaluations. In the following, we
will discuss relevant challenges and surprising findings from our
workshop, including the on-boarding process, the importance of the
environment design, the influence of users' hardware, and the novel
interactions made possible by virtual environments. We conclude with
some design insights to inspire future work.


Novel experiences always face on-boarding challenges. These challenges
typically include usability, sociability, and learning curve
components. The latter tends to be a product of the first two. During
the interviews, participants discussed how the virtual workshop was
similar to a ``first conference'' experience. In this case, the lack
of familiarity with the environment might have been an advantage.
Many participants had their first experiences with Mozilla Hubs during
our on-boarding process, which established the social norms of an
academic workshop.  VEs are frequently used for events like DJ parties
or art museum experiences, but one of our main objectives was to
create a conference experience without feeling game-like or drawing
from gaming metaphors. For participants familiar with game platforms
like Minecaft or World of Warcraft, developing new norms and standards
for non-game experiences could've been an additional challenge. The
on-boarding process was positively received and like any other novel
interface, it was a vital part of the virtual workshop
experience. It's advisable to have both a written short introduction
to the hosting platform and a short pre-workshop session on the
platform in order to familiarise participants with the environment and
expected social norms~\cite{Le2020EnhancingEnvironments}.

Mozilla Hubs currently does not provide much \textit{social
  translucence}~\cite{Erickson2000}, which is most problematic when
moving between different linked environments.  Social translucence, or
more simply social awareness, states digital system design should
support coherent behaviour by making actions visible to all
participants. In a simple example, social cues or other observed
behaviours should be communicated---like User B left the room or User
Z is typing---to maintain visibility, awareness, and accountability.
While our environment contained portals linking the breakout rooms to
the main room, there was no visibility \textit{between} rooms.  For
example, the only way to see if someone is in a different room is to
travel there~\cite{Le2020EnhancingEnvironments}.  This can be a
disruptive experience, which requires one to fully exit one space
before gaining awareness of the state of the new space.  For
observers, someone entering a portal instantly disappears from the
current room as if they lost connectivity or closed their browser
window.  These portals are uni-directional and provide limited
information about the destination, for example if one will be able to
easily return if they use the portal.  Increased social translucence
must provide feedback that decreases the disruption of moving between
spaces, gives awareness of the state of destination, and makes
movements between spaces visible. 

In guided virtual workshops, such as
ours, translucency could provide a way for organisers to ensure all
attendants are stationed in the rooms they are meant to be at any
given time. In their paper, Lu et
al.~\cite{Le2020EnhancingEnvironments} talk of different motivations
to attend a conference. Ultimately, all attendees want to learn about
new research findings and connect with others. However, remote
attendees wanted to connect with new people while local attendees were
interested in socialising with existing or potential collaborators
they already knew about. Initialising a conversation is made more
difficult especially if you are a remote attendee with no easily
recognisable contacts as implied by some participants (P3, P9). In the
future, it would be interesting to investigate how to allow attendees
and moderators to have a better awareness of who else is attending,
which virtual room they are in and also enable them to capture,
interpret and express the right social cue to turn-take and indicate
interest in joining a group conversation regardless of which device
the attendee is using. In our workshop, we saw some unintended but
delightful interactions and behaviours displayed around the `tiki
corner' and the `ocean' outside the breakout room. It appears that
even in serious fit-for-purpose virtual space, there is some mileage
in including what may seem as superfluous objects in order to
encourage ice-breaking social behaviour---jumping into the ocean with
strangers or taking group selfies with friends.

Based on our quantitative proxemic dataset, we learned about the
influence that a VE can have on the participants and their
behaviours. The \textit{Outdoor Meetup} space included a large screen,
a podium and a mini-theatre, intended for presentations. The keynote
speaker decided instead to project his slides to the back of the room,
making the reserved keynote space irrelevant.  The ability to
dynamically reconfigure space easily in a VE allows for unexpected
space occupancy and group formations.  Dynamic changes like this can
change the way people use virtual space in way unanticipated by the
environment's designers.  While some designed features proved
irrelevant, some incidental features had unexpected impacts.  For
example, the floor pattern rendered as a texture on the floor plane
had a strong influence on the positioning of the participants, as
shown in Figure~\ref{fig:pos_overview}. Participants avoided the floor
texture and appeared to surf the edge when moving in the space. These
insights open new areas of research around data-driven design of VEs.
For example, can we gather and process large-scale proxemic and
movement data in order to automatically generate spaces for different
purposes and influence pedestrian flow and social grouping? Can the
environments self-adapt based on the actual use in order to increase
engagement? Mathematical approaches to space
design~\cite{Hillier1988TheSpace} could have new relevance in VEs when
proxemic data is continuously available.

As a 3D virtual environment, spatial relationships form a large part
of the user experience in Mozilla Hubs.  The interface provides a
field of view, and users must navigate the VE and interact with others
through this view.  Our participants were aware of spatial
relationships and spread out during the workshop. The amount of
personal space participants maintained depended on the activity and
the design of the environment. Although, we introduced participants to
Mozilla Hubs and its functionality, there were some virtual skills we
didn't cover in that initial introduction. However, similar to in real
world situations, participants could observe others' actions, like
flying, and would ask each other how they too might fly. Similar to
other reports~\cite{Le2020EnhancingEnvironments}, flying was seen as a
fun activity but it wasn't always deemed ``acceptable,'' as we only
observed flying regularly during the keynote. This `stadium seating
pattern' was not observed during other kinds of activities like small
breakout groups. The \textit{Lake Office} rooms may have created a
sense of ``perceived containment'' from incorrectly assuming walls
would act as barriers or a social expectation to remain visible within
the walls.  Flying in the larger \textit{Outdoor Meetup} space was
also rare during informal breaks. Social expectations may discourage
such behaviour that could appear like aggressively seeking privacy.

Previous works~\cite{Schroeder2001, Slater2000, Steed1999} demonstrate
that asymmetry between participants on different systems affects their
collaboration in a task. Participants with a network disadvantage, for
instance, were unlikely to emerge as the `leader' in a collaborative
task and participants using an HMD were more likely to emerge as the
`leader.' In our workshop, in order to handle this, we had assigned
moderators who were the automatic guides regardless of device
capability. Still, the experiences of users with HMDs and users
without HMDs were significantly different. In Mozilla Hubs,
participants using HMDs have hands rendered as part of their avatar,
while participants using a web browser do not. This allowed the HMD
users to be more expressive and gesture. Having hands in the VE aided
casual conversation and allowed HMD users to add additional
information to their speech, for example using diectic gestures during
their presentation. This was observed in other virtual events as
well~\cite{Le2020EnhancingEnvironments}, where audiences watching a
presentation wanted to applaud at the end. Although an imbalance of
expression was noticeable, there were also some advantages when using
an HMD\@. HMD users could not share a screen or a document as easily
as the moderators in the fast paced breakout sessions. Finally,
participants using a browser on a computer or a mobile device have a
more standard `video call' feeling as they could move the window aside
while looking for a document or taking notes. The moderators in all
breakout rooms chose to attend the workshop using their desktop in
order to facilitate and take live notes. This allowed for a relatively
long but very interactive workshop---half a working day. The lesson we
learned regarding the influence of the local configuration is the need
for adequate customisation or intervention elements to minimise their
imbalanced experiences. In the future, it would be interesting to
study groups of workshop attendees where the role of a moderator is
undefined. For instance, does the note-taker and facilitator roles in
a small group always fall upon the one attendee on a desktop if
everyone else is attending using an HMD, even though at first glance
it might seem as if the desktop is not the most advanced device to use
when attending virtual workshops?

\section{Conclusion}
Remote events can benefit from virtual environments. In
the context of an academic workshop, we found that proxemic interactions are
congruent with those in the physical, although VEs are not constrained
by physics and we observed that flying could add a new dimension to
personal space.  Our results demonstrate how group formations adapt to
the world/room size and the ability to create conversation clusters.
Further our approach can provide novel insights into the effect of the environment
textures, shapes, and routing.  While some social cues were analogous
to real world interactions, there is a need for expression,
indicators, and translucence when using a generic VE like Mozilla
Hubs. Being open source, Mozilla Hubs is extensible to gain precise
user actions and position data at tick resolution to analyse and
answer questions on social behaviours as well as providing ample data
for future insights and possible generative layouts to support
collaborative meetings and tasks.

\begin{acks}
  The authors would like to thank to Yulius Tjahjadi for standing up
  our Mozilla Hubs Cloud beta instance in AWS and Blair MacIntyre at
  Georgia Tech for keynoting the workshop.  We extend additional
  thanks to Liv, Elgin, and the rest of Mozilla Hubs team for their
  assistance in setting up the VE\@.  We also thank all the workshop
  participants for joining us during this event and making this
  research possible.
  
  This research was partially funded by the European Commission as
  part of the H2020 program under the grant agreement \#762111
  ``VRTogether''~\url{http://vrtogether.eu/} (accessed January 2021).
  
  This research was partially funded by the European Commission as
  part of the H2020 ERC project ViAjeRo, grant agreement
  \#835197.~\url{https://viajero-project.org/} (accessed January
  2021).
\end{acks}

\bibliographystyle{ACM-Reference-Format}
\bibliography{unified-references}


\appendix
\section{Supplemental Materials}\label{app:dataset}
The supplemental materials with this article include:
\begin{itemize}
\item JavaScript code to instrument Mozilla Hubs Cloud for data
  logging.  Two of the detailed instrumented
  items,~\ref{sec:instrument}\ref{sec:instr-virt-envir:talking}
  and~\ref{sec:instrument}\ref{sec:instr-virt-envir:damp}, were later
  additions to the collection code and not available during our study.
\item The anonymous dataset collected from our workshop.
\item A collection of Python Jupyter Notebooks to clean and analyse
  the data, generate the figures and animation in this paper.
\end{itemize}  
These materials are available in the ACM Digital Library, as well as, 
on Github at \url{https://github.com/ayman/hubs-research-acm-chi-2021} 
(accessed January 2021).
\end{document}